\documentclass[aps,prd,twocolumn,showpacs,amsmath,amssymb]{revtex4-1}
\usepackage{amsmath}
\usepackage{graphicx}
\usepackage{subfigure}
\usepackage{epstopdf}
\usepackage{color}
\usepackage{multirow}
\usepackage{setspace}
\usepackage{overpic}
\usepackage{amssymb}
\usepackage[bookmarksnumbered, pdfstartview=FitH,colorlinks,urlcolor=blue, citecolor=blue,linkcolor=blue] {hyperref}
\usepackage{lineno}
\usepackage{bm}
\usepackage{rotating}
\usepackage[utf8]{inputenc}
\usepackage{makecell}

\let\oldequation\equation
\let\oldendequation\endequation

\renewenvironment{equation}
  {\linenomathNonumbers\oldequation}
  {\oldendequation\endlinenomath}

\begin{document}

\title{\bf \boldmath
Study of the doubly Cabibbo-suppressed decays $D^+_s\to K^+K^+\pi^-$ and $D^+_s\to K^+K^+\pi^-\pi^0$}

\author{
M.~Ablikim$^{1}$, M.~N.~Achasov$^{4,b}$, P.~Adlarson$^{75}$, X.~C.~Ai$^{81}$, R.~Aliberti$^{35}$, A.~Amoroso$^{74A,74C}$, M.~R.~An$^{39}$, Q.~An$^{71,58}$, Y.~Bai$^{57}$, O.~Bakina$^{36}$
% ...
\\
\vspace{0.2cm}
(BESIII Collaboration)\\
\vspace{0.2cm} {\it
$^{1}$ Institute of High Energy Physics, Beijing 100049, People's Republic of China\\
% ...
$^{4}$ Budker Institute of Nuclear Physics SB RAS (BINP), Novosibirsk 630090, Russia\\
% ...
$^{35}$ Johannes Gutenberg University of Mainz, Johann-Joachim-Becher-Weg 45, D-55099 Mainz, Germany\\
$^{36}$ Joint Institute for Nuclear Research, 141980 Dubna, Moscow region, Russia\\
% ...
$^{39}$ Liaoning Normal University, Dalian 116029, People's Republic of China\\
% ...
$^{57}$ Southeast University, Nanjing 211100, People's Republic of China\\
$^{58}$ State Key Laboratory of Particle Detection and Electronics, Beijing 100049, Hefei 230026, People's Republic of China\\
% ...
$^{71}$ University of Science and Technology of China, Hefei 230026, People's Republic of China\\
% ...
$^{74}$ University of Turin and INFN, (A)University of Turin, I-10125, Turin, Italy; (B)University of Eastern Piedmont, I-15121, Alessandria, Italy; (C)INFN, I-10125, Turin, Italy\\
$^{75}$ Uppsala University, Box 516, SE-75120 Uppsala, Sweden\\
% ...
$^{81}$ Zhengzhou University, Zhengzhou 450001, People's Republic of China\\
\vspace{0.2cm}
% ...
$^{b}$ Also at the Novosibirsk State University, Novosibirsk, 630090, Russia\\
% ...
}
}

\vspace{100 pt}
 \date{\today}

\begin{abstract}
Based on 7.33 fb$^{-1}$  of $e^+e^-$ collision data collected at center-of-mass energies between 4.128 and 4.226 GeV with the BESIII detector, the experimental studies of the doubly Cabibbo-suppressed  decays $D^+_s\to K^+K^+\pi^-$ and $D^+_s\to K^+K^+\pi^-\pi^0$ are reported.
We determine the absolute branching fraction of $D^+_s\to K^+K^+\pi^-$ to be (${1.23^{+0.28}_{-0.25}}({\rm stat})\pm0.06({\rm syst})$) $\times 10^{-4}$.
No significant signal of $D^+_s\to K^+K^+\pi^-\pi^0$ is observed and
the upper limit on its decay branching fraction at 90\% confidence level is set to be $1.7\times10^{-4}$.
\end{abstract}

\maketitle

\oddsidemargin  -0.2cm
\evensidemargin -0.2cm

\section{Introduction}

Doubly Cabibbo-suppressed (DCS) decays of charmed mesons offer a unique platform to understand the dynamics of charmed mesons.
It is naively expected that the ratio of the branching fraction between a given DCS $D^{0(+)}$ decay and its Cabibbo-favored (CF) counterpart
is about $(0.5-2.0)\times {\rm tan}^4\theta_C$,
where $\theta_C$ is the Cabibbo mixing angle~\cite{Lipkin,theory_1}.
In 2020 and 2021, BESIII reported \cite{bes3_DCS_Kpipipi0,bes3-DCS-Dp-K3pi-v2} the observation of the DCS
decay $D^+\to K^+\pi^+\pi^-\pi^0$ (charge conjugate decays are always implied throughout this paper).
The average branching fraction of $D^+\to K^+\pi^+\pi^-\pi^0$, weighted from the two measurements in~\cite{bes3_DCS_Kpipipi0,bes3-DCS-Dp-K3pi-v2}, is
$[1.13 \pm 0.08({\rm stat}) \pm 0.03({\rm syst})]\times 10^{-3}$.
This gives a DCS/CF branching fraction ratio of $(6.3\pm0.5)\tan^4\theta_C$.
Further measurements of the DCS decays of charmed mesons may shed light on the decay dynamics.  Specifically, two-body DCS $D$ decay branching fractions are important inputs to help understand quark SU(3)-flavor symmetry and its breaking effects in the charm sector~\cite{pap4,pap5,pap6,pap7,pap8,pap9,pap10}.

In the $D^+_s$ sector, only the DCS decay $D^+_s\to K^+K^+\pi^-$ was previously reported~\cite{pdg2022}.
Its decay branching fraction was measured relative to the CF decay of $D^+_s\to K^+K^-\pi^+$ by LHCb, BaBar, Belle and
FOCUS~\cite{pap34,pap35,pap36,pap37}.
In this article, we present the first measurement of absolute branching fraction of $D_s^+\to K^+K^+\pi^-$ and search for $D_s^+\to K^+K^+\pi^-\pi^0$ for the first time.
This analysis uses the $e^+e^-$ collision data samples
collected at center-of-mass energies ($E_{\rm cm}$) between 4.128 and 4.226 GeV with the BESIII detector,
corresponding to an integrated luminosity of 7.33 fb$^{-1}$.

\section{BESIII detector and Monte Carlo simulation}

The BESIII detector \cite{BESIII} is a magnetic
spectrometer located at the Beijing Electron
Positron Collider (BEPCII)~\cite{Yu:IPAC2016-TUYA01}. The
cylindrical core of the BESIII detector consists of a helium-based
 multilayer drift chamber (MDC), a plastic scintillator time-of-flight
system (TOF), and a CsI(Tl) electromagnetic calorimeter (EMC),
which are all enclosed in a superconducting solenoidal magnet
providing a 1.0~T magnetic field. The solenoid is supported by an
octagonal flux-return yoke with resistive plate counter muon
identifier modules interleaved with steel. The acceptance of
charged particles and photons is 93\% over $4\pi$ solid angle. The
charged particle momentum resolution at $1~{\rm GeV}/c$ is
$0.5\%$, and the specific ionization energy loss $dE/dx$ resolution is $6\%$ for the electrons
from Bhabha scattering. The EMC measures photon energies with a
resolution of $2.5\%$ ($5\%$) at $1$~GeV in the barrel (end cap)
region. The time resolution of the TOF barrel section is 68~ps,
while that of the end cap portion is 110~ps. The end cap TOF system was upgraded in 2015 using multi-gap resistive plate chamber technology, providing a time resolution of 60 ps~\cite{Pcao}, which benefits 83\% of the data used in this analysis~\cite{dt1,dt2}.

Simulated samples produced with the {\sc geant4}-based~\cite{geant4} Monte-Carlo (MC) package which
includes the geometric description of the BESIII detector and the
detector response, are used to determine the detection efficiency
and to estimate the backgrounds. The simulation includes the beam
energy spread and initial state radiation (ISR) of the $e^+e^-$
annihilations modeled with the generator {\sc kkmc}~\cite{kkmc}.
The inclusive MC samples consist of the production
of $D_{s}^{(*)}\bar{D_{(s)}^{(*)}}$ pairs
(with consideration of quantum coherence for all $D^0\bar{D}^0$
pair decays), the non-$D\bar{D}$ decays of the $\psi(3770)$, the ISR
production of the $J/\psi$ and $\psi(3686)$ states, and the
continuum processes.
The known decay modes are modeled with {\sc
evtgen}~\cite{evtgen} using the branching fractions taken from the
Particle Data Group (PDG)~\cite{pdg2022}, and the remaining unknown decays
of the charmonium states with {\sc lundcharm}~\cite{lundcharm}.
Final state radiation
from charged final state particles is incorporated with the {\sc
photos} package~\cite{photos}.
The signal MC sample of $D^+_s\to K^+K^+\pi^-$ is generated
using the known fraction of the intermediate channel $D^+_s\to K^+K^{*0}$ \cite{pdg2022},
while the unknown sub-channels are generated according to the phase space.
The signal MC sample of $D^+_s\to K^+K^+\pi^-\pi^0$ is also generated uniformly over the phase space.

\section{Measurement method}

In the $e^+e^-$ collision data taken at $E_{\rm cm}$ between 4.128 and 4.226~GeV,
the $D_s^\pm$ mesons are produced mainly via the $e^+e^-\to D_s^{*\pm}D_s^\mp\to \gamma(\pi^0)D_s^+ D_s^-$ process.
This analysis is performed by using the double-tag (DT) method pioneered by the MARKIII Collaboration~\cite{DTmethod}.
A $D_s^-$ meson which is fully reconstructed via one of the eleven hadronic decay modes
%
%
is referred to as a single-tag (ST) $D_s^-$ meson.
The event, in which the $\gamma(\pi^0)$ emitted from $D_s^{*+}$ and the signal decay can be successfully
reconstructed in the presence of ST $D^-_s$ meson, is called as a DT event.
The branching fraction of the signal decay is determined as
\begin{equation}
\mathcal B_{\rm sig}=\frac{N_{\rm DT}}{N_{\rm ST} \cdot \epsilon_{\rm sig}},
\label{eq1}
\end{equation}
where $N_{\rm DT}$ and $N_{\rm ST}$ are the yields of the DT events and ST $D^-_s$ mesons in data, respectively;
$\epsilon_{\rm sig}$ is the efficiency of detecting the signal decay in the presence of the ST $D^-_s$ mesons, averaged over the $D_s^* \to D_s\gamma$ and $D_s^* \to D_s\pi^0$ transitions weighted by their branching fractions.

\section{ST candidates}

To reconstruct ST $D^-_s$ candidates, we use eleven hadronic decay modes of
$D^-_s\to K^{+} K^{-}\pi^{-}$,
$K^{+} K^{-}\pi^{-}\pi^{0}$,
$\pi^{-}\pi^{+}\pi^{-}$,
$K_S^{0} K^{-}$,
$K_S^{0} K^{+}\pi^{-}\pi^{-}$,
$\eta_{\gamma\gamma}\pi^{-}$,
$\eta_{\pi^{+}\pi^{-}\pi^{0}}\pi^{-}$,
$\eta\prime_{\pi^{+}\pi^{-}\eta} \pi^{-}$,
$\eta\prime_{\gamma\rho^{0}} \pi^{-}$,
$\eta_{\gamma\gamma}\rho^{-}$, and
$\eta_{\pi^{+}\pi^{-}\pi^{0}}\rho^{-}$.
Throughout this article, $\rho$ denotes $\rho(770)$ and the subscripts of $\eta^{(\prime)}$ denote individual decay modes adopted for $\eta^{(\prime)}$ reconstruction.

\begin{figure}[htbp]
\centering
\includegraphics[width=0.5\textwidth] {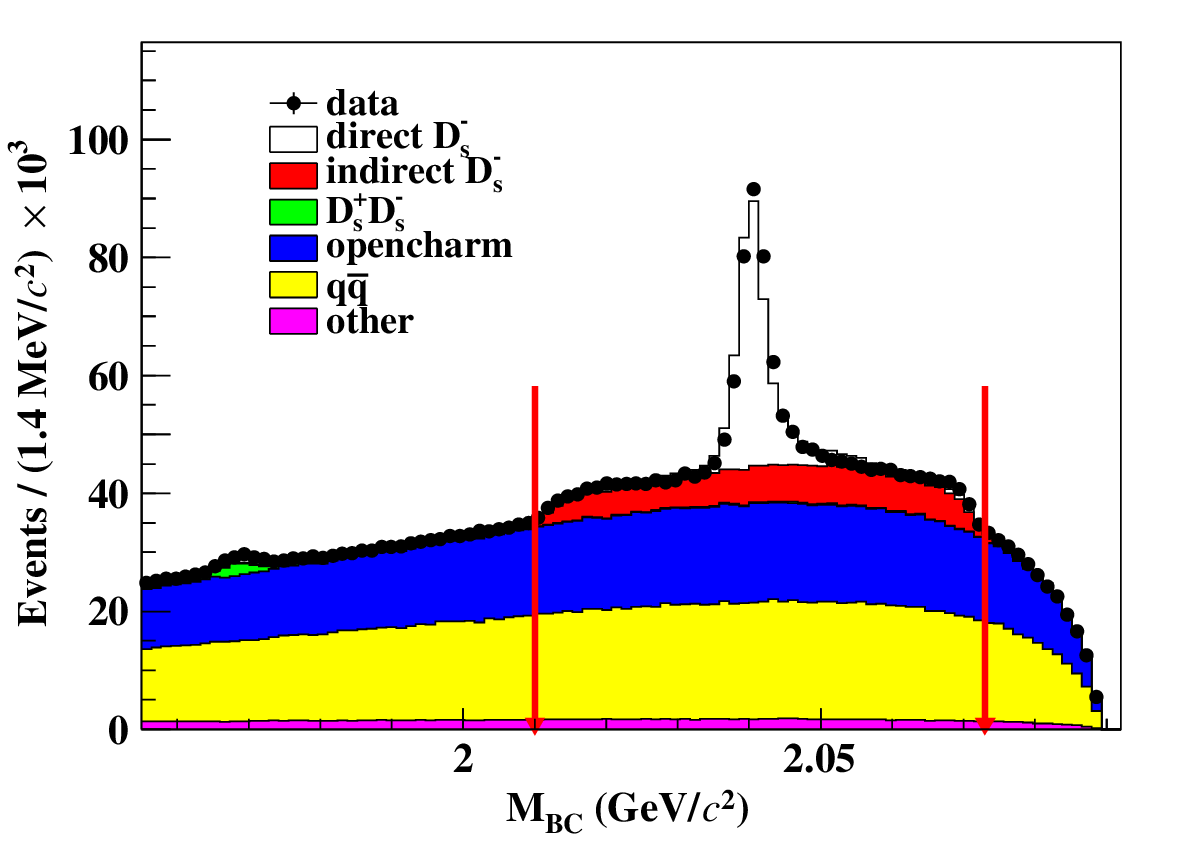}
\caption{
The $M_{\rm BC}$ distributions of the ST candidates in data and inclusive MC samples at 4.178 GeV. The candidates between the two red arrows are accepted.
}
\label{fig:mBC}
\end{figure}

\begin{table}[htbp]
	\centering\linespread{1.15}
	\caption{Requirements of $M_{\rm BC}$ for each energy point.}
	\small
	\label{tab:mbc}
	\begin{tabular}{cc}
		\hline\hline
		$E_{\rm cm}$ (GeV) & $M_{\rm BC}$ (GeV/$c^2$) \\
		\hline
		4.128          & $[2.010,2.061]$         \\
		4.157          & $[2.010,2.070]$         \\
		4.178          & $[2.010,2.073]$         \\
		4.189          & $[2.010,2.076]$         \\
		4.199          & $[2.010,2.079]$         \\
		4.209          & $[2.010,2.082]$         \\
		4.219          & $[2.010,2.085]$         \\
		4.226          & $[2.010,2.088]$         \\
		\hline\hline
	\end{tabular}
\end{table}

The $K^\pm$, $\pi^\pm$, $K^0_S$, $\gamma$, $\pi^0$, and $\eta$ candidates are selected with the same criteria as in Refs.~\cite{bes2019,bes3_etaev,bes3_gev}.
All charged tracks, except for those from $K^0_S$, are required to originate from a region defined as
 $|V_{xy}|<1$~cm and $|V_{z}|<10$~cm,
where $|V_{z}|$ and $|V_{xy}|$ are the distances of the closest approach
relative to the interaction
point along the MDC axis and in the transverse plane, respectively.
The track polar angle $\theta$ with respect to the MDC axis
must satisfy $|\!\cos\theta|<0.93$.
Charged particles are identified by using the combined $dE/dx$ and TOF information.
A particle is assigned to be a pion (kaon) candidate if the confidence level
for the corresponding hypothesis is greater than for the kaon (pion) hypothesis.

Candidates for $K_S^0$ are selected via the decay $K^0_S\to \pi^+\pi^-$.
The two charged pions are required to satisfy $|V_{z}|<20$ cm and $|\!\cos\theta|<0.93$.
They are assumed to be $\pi^+\pi^-$ without particle identification (PID) requirements. The two pions are required to have a common vertex point via a secondary vertex fit; the fit $\chi^2$ must be less than 200 and their invariant mass is required to be within $(0.487, 0.511)$ GeV$/c^2$.
The decay length of any $K_S^0$ candidate, measured from the interaction
point, is required to be greater than twice the vertex resolution.

Photon candidates are reconstructed by using shower information measured
by the EMC.  To suppress backgrounds from electronic noise or bremsstrahlung,
candidate showers are required to start within [0, 700] ns from the event
start time and the energy of each shower in the barrel (end cap) region
of the EMC~\cite{BESIII} is required to be greater than 25 (50) MeV.
To suppress backgrounds associated with charged tracks,
the angle subtended by the EMC shower and the closest charged track
extrapolation to the EMC must be greater than 10 degrees as measured from the interaction point.

\begin{figure*}[htbp]
\centering
\includegraphics[width=1.0\textwidth]{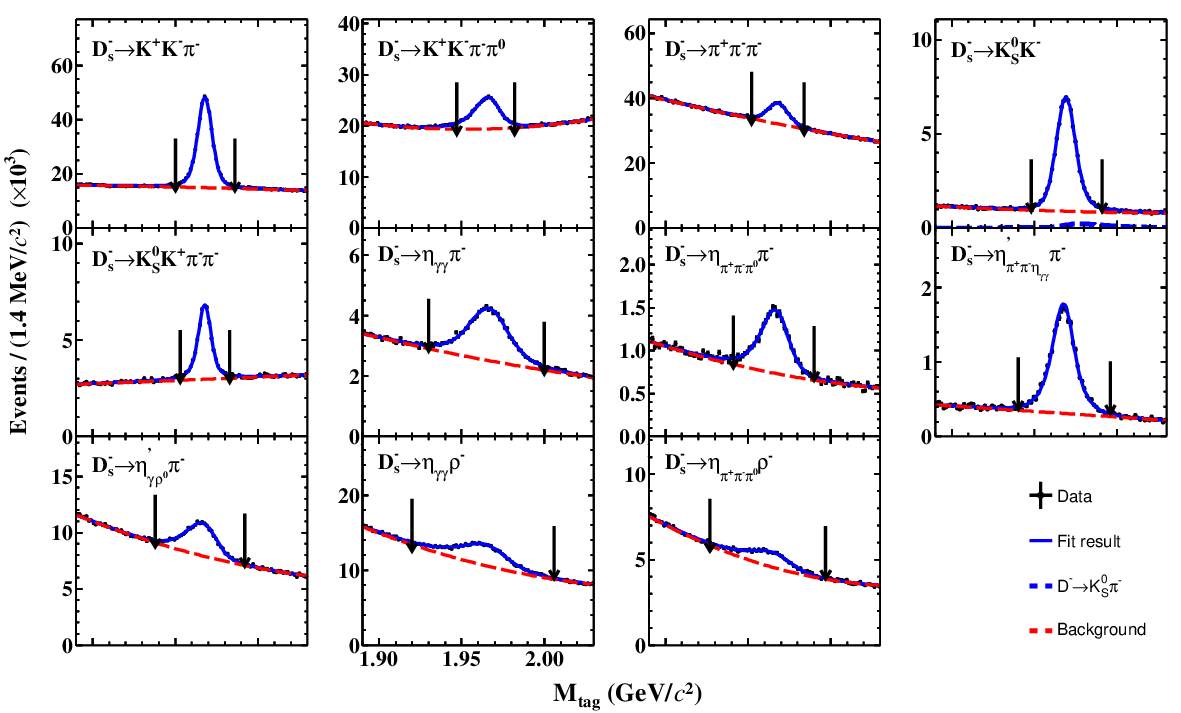}
\caption{\footnotesize
Fits to the $M_{\rm tag}$ distributions of the selected ST candidates for various tag modes.
The points with error bars are the data combined from all energy points.
The blue solid curves represent the best fit results,
and the red dashed curves stand for the fitted backgrounds.
For the tag mode of $D^-_s\to K_S^0K^-$, the blue dashed curve is the peaking background from $D^-\to K_S^0\pi^-$.
The pair of arrows denote the $M_{\rm tag}$ signal regions.
}
\label{fig:stfit}
\end{figure*}

Candidates for $\pi^0$ and $\eta_{\gamma\gamma}$ are formed from $\gamma\gamma$ pairs with invariant masses in the intervals $(0.115,0.150)$ and $(0.500,0.570)$~GeV$/c^{2}$, respectively.
To improve momentum resolution, each $\gamma\gamma$ pair is subject to a one-constraint (1C) kinematic fit that constrains their invariant mass to the $\pi^{0}$ or $\eta$ nominal mass~\cite{pdg2022}. To form candidates for $\rho^{+(0)}$, $\eta_{\pi^0\pi^+\pi^-}$, $\eta^\prime_{\eta\pi^+\pi^-}$, and $\eta^\prime_{\gamma\rho^0}$,
the invariant masses of the $\pi^+\pi^{0(-)}$, $\pi^0\pi^+\pi^-$, $\eta\pi^+\pi^-$, and  $\gamma\rho^0$ combinations are required to be
within the mass intervals of  $(0.570,0.970)$, $(0.530,0.570)$, $(0.946,0.970)$ and $(0.940,0.976)~\mathrm{GeV}/c^2$, respectively.
In addition, the energy of the $\gamma$ from an $\eta^\prime_{\gamma\rho^0}$ decay is required to be greater than 0.1~GeV.

The transition pions from $D^{*+}$ decays are suppressed by requiring the momentum of any pion which is not from a $K_S^0$, $\eta$, or $\eta^\prime$ decay to be greater than 0.1~GeV/$c$. In order to reject peaking background from the $D^-_s\to K^0_S\pi^-$ final state in the selection of the $D^-_s\to \pi^+\pi^-\pi^-$ tag mode, the invariant mass of each $\pi^+\pi^-$ combination is required to be outside the mass window of $(0.468, 0.528)$ GeV/$c^2$.

The backgrounds from non-$D_s^{\pm}D^{*\mp}_s$ processes are suppressed by using the beam-constrained mass of the ST $D_s^-$ candidate,
defined as
\begin{equation}
M_{\rm BC}\equiv\sqrt{E^2_{\rm beam}-|\vec{p}_{\rm tag}|^2},
\end{equation}
where $E_{\rm beam}$ is the beam energy and
$\vec{p}_{\rm tag}$ is the momentum of the ST $D_s^-$ candidate in the $e^+e^-$ rest frame. Figure~\ref{fig:mBC} shows the $M_{\rm BC}$ distributions of the ST candidates in data and inclusive MC samples at 4.178 GeV. For the direct $D^-_s$ mesons  which are produced by the $e^+e^-$ collision,
$M_{\rm BC}$ peaks, as shown by the open histogram.
For the indirect $D^-_s$ mesons  which are produced from $D^{*-}_s$ decays,
the $M_{\rm BC}$ distribution is broader, as shown by the red histogram.
The $M_{\rm BC}$ value is required to be within the intervals listed in Table~\ref{tab:mbc}.
This requirement retains about 99\% of the $D_s^-$ and $D_s^+$ mesons from $e^+ e^- \to D_s^{*\mp}D_s^{\pm}$.

If there are multiple candidates present  per tag mode per charge,
only the one with the $D_s^-$ recoil mass
\begin{equation}
M_{\rm rec} \equiv \sqrt{ \left (E_{\rm cm} - \sqrt{|\vec p_{\rm tag}|^2+m^2_{D^-_s}} \right )^2
-|\vec p_{\rm tag}|^2}
\end{equation}
closest to the $D_s^{*+}$ nominal mass~\cite{pdg2022} is kept for further analysis.  When the $D_s^-$ tag meson is from a $D^{*-}$ decay, the recoil
mass distribution is not sharply peaked, but overall the correct tag is selected about 94\% of the time.
The distributions of the invariant masses ($M_{\rm tag}$) of the accepted ST candidates for various tag modes are shown in Fig.~\ref{fig:stfit}.
The yields of ST $D^-_s$ mesons reconstructed in various tag modes are derived from fits to their individual $M_{\rm tag}$ distributions.
In the fits, the signal is described by the simulated shape convolved with a Gaussian function to take into account the resolution difference between data and simulation.
In the fit to the $D_s^-\to K_S^0K^-$ tag mode,
the shape of the peaking background of $D^-\to K^0_S\pi^-$ is modeled by the simulated shape convolved with the same Gaussian resolution function as used for the signal.
The combinatorial background is described by first, second or third-order Chebychev polynomial functions. The order of the polynomial is chosen by analyzing the inclusive MC sample.
Figure~\ref{fig:stfit} shows the fit results of different tag modes for the combined data sample from all energy points.
In each sub-figure, the black arrows show the chosen $M_{\rm tag}$ signal regions.
The candidates located in these signal regions are kept for further analysis.

Based on simulation, the $e^+e^-\to(\gamma_{\rm ISR})D_s^+D_s^-$ process is found to contribute about (0.7-1.1)\% in the fitted yields of ST $D^-_s$ mesons for various tag modes and has been subtracted away from the fitted yields in this analysis.  Production of $D_s^{*+}D_s^{*-}$ pairs occurs only at the highest c.m.~energy; candidates are suppressed by the $M_{\rm BC}$ requirement and are negligible.
The second and third columns of Table~\ref{tab:bf} summarize the yields of ST $D^-_s$ mesons ($N_{\rm ST}$) for various tag modes obtained from
the combined data sample and the corresponding detection efficiencies ($\epsilon_{\rm ST}$), respectively.

\begin{table*}[htbp]
\centering\linespread{1.2}
\caption{
The fitted yields of ST $D^-_s$ mesons from the combined data sample, $N_{\rm ST}$ (in units of $\times 10^3$),
the efficiencies of detecting ST $D^-_s$ mesons, DT events and signal events ($\epsilon_{\rm ST}$, $\epsilon_{\rm DT}$, $\epsilon_{\rm sig}=\epsilon_{\rm DT}/\epsilon_{\rm ST}$, all \%) for various tag modes.
For all numbers, the uncertainties are statistical only.
The listed efficiencies do not include the branching fractions of any daughter particle decays.
For each signal channel, the efficiencies have been weighted over different energy points. }
\centering
\small
        \label{tab:bf}
        \begin{tabular}{l |p{2.1cm}<{\centering} |p{1.8cm}<{\centering}  p{1.8cm}<{\centering}|p{2.2cm}<{\centering} p{2.2cm}<{\centering} |p{2.3cm}<{\centering} p{2.3cm}<{\centering}}\hline\hline
Tag mode &{${\rm M}_{\rm tag}({\rm GeV}/c^2)$} &$N_{\rm ST}$ &$\epsilon_{\rm ST}$ &$\epsilon_{{\rm DT},K^+K^+\pi^-}$ &$\epsilon_{{\rm sig},K^+K^+\pi^-}$ &$\epsilon_{{\rm DT},K^+K^+\pi^-\pi^0}$ &$\epsilon_{{\rm sig},K^+K^+\pi^-\pi^0}$\\ \hline
$K^{+} K^{-}\pi^{-}$                            &(1.950,1.986)                         &280.7$\pm$0.9 &40.87$\pm$0.01  &13.99$\pm$0.06 &34.23$\pm$0.14  &4.24$\pm$0.03  &10.38$\pm$0.09\\
$K^{+} K^{-}\pi^{-}\pi^{0}$                     &(1.947,1.982)       &86.3$\pm$1.3  &11.83$\pm$0.01  &4.53$\pm$0.09  &38.30$\pm$0.80  &1.08$\pm$0.05  &9.16$\pm$0.44 \\
$\pi^{-}\pi^{+}\pi^{-}$                         &(1.952,1.984)       &72.7$\pm$1.4  &51.86$\pm$0.03  &17.68$\pm$0.06 &34.09$\pm$0.12  &5.54$\pm$0.04  &10.69$\pm$0.07\\
$K_S^{0} K^{-}$                                 &(1.948,1.991)       &62.2$\pm$0.4  &47.37$\pm$0.03  &16.07$\pm$0.07 &33.93$\pm$0.15  &5.07$\pm$0.04  &10.70$\pm$0.09\\
$K_S^{0} K^{+}\pi^{-}\pi^{-}$                   &(1.953,1.983)       &29.6$\pm$0.3  &20.98$\pm$0.03  &6.99$\pm$0.14  &33.31$\pm$0.68  &1.85$\pm$0.08  &8.82$\pm$0.38 \\
$\eta_{\gamma\gamma}\pi^{-}$                    &(1.930,2.000)       &39.6$\pm$0.8  &48.31$\pm$0.04  &16.79$\pm$0.06 &34.76$\pm$0.12  &5.19$\pm$0.03  &10.74$\pm$0.07\\
$\eta_{\pi^{+}\pi^{-}\pi^{0}}\pi^{-}$           &(1.941,1.990)       &11.7$\pm$0.3  &23.31$\pm$0.05  &8.21$\pm$0.06  &35.24$\pm$0.26  &2.35$\pm$0.03  &10.07$\pm$0.15 \\
$\eta\prime_{\pi^{+}\pi^{-}\eta} \pi^{-}$       &(1.940,1.996)       &19.7$\pm$0.2  &25.17$\pm$0.04  &8.32$\pm$0.06  &33.06$\pm$0.23  &2.46$\pm$0.03  &9.78$\pm$0.13 \\
$\eta\prime_{\gamma\rho^{0}} \pi^{-}$           &(1.938,1.992)       &50.4$\pm$1.0  &32.46$\pm$0.03  &11.53$\pm$0.06 &35.51$\pm$0.18  &3.42$\pm$0.03  &10.52$\pm$0.11\\
$\eta_{\gamma\gamma}\rho^{-}$  &(1.920,2.006)       &80.1$\pm$2.3  &19.92$\pm$0.01  &8.31$\pm$0.07  &41.72$\pm$0.30  &2.27$\pm$0.04  &11.40$\pm$0.18 \\
$\eta_{\pi^{+}\pi^{-}\pi^{0}}\rho^{-}$      &(1.927,1.997)    &22.2$\pm$1.4  &9.15$\pm$0.01   &3.90$\pm$0.07  &42.64$\pm$0.65  &0.98$\pm$0.03  &10.73$\pm$0.38 \\
Weighted Ave.                                         &      &              &                &               &35.57$\pm$0.18  &               &10.32$\pm$0.10 \\
\hline\hline
\end{tabular}
\end{table*}

\section{DT candidates}

The transition photon (or $\pi^0$) and the signal $D^+_s$ decay candidate
are reconstructed from the particles recoiling against the selected
$D_s^-$ tag.  We define the energy difference
$\Delta E \equiv  E_{\rm tag} + E_{\gamma(\pi^0)+D^-_s}^{\rm rec} + E_{\gamma(\pi^0)} - E_{\rm cm}$,
where
$E_{\gamma(\pi^0)+D^-_s}^{\rm rec} \equiv \sqrt{|-\vec{p}_{\gamma(\pi^0)}-\vec{p}_{\rm tag}|^2 +
  M^2_{D_s^+}}$, $E_i$ and $\vec{p}_i$ [$i = \gamma(\pi^0)$ or tag] are the energy and momentum of
$\gamma(\pi^0)$ or $D_s^-$ tag, respectively.
A loop is performed over all $\gamma$ and $\pi^0$ candidates which were not used in the ST reconstruction, under the two assumptions of $D^*_s\to \pi^0D_s$ and $D^{*+}_s\to \gamma D^+_s$. If more than one combination satisfies the selection criteria, the one with the minimum $|\Delta E|$ is chosen.

The selection criteria of $\pi^-$, $K^+$, and $\pi^0$ for
$D^+_s\to K^+K^+\pi^-$ and $D^+_s\to K^+K^+\pi^-\pi^0$ reconstruction
are the same as those used to choose the ST candidates.
We require that there are exactly three good charged tracks which are not used to form the ST candidate; one track is identified as a $\pi^-$
and the others are required to be identified as $K^+$s.
For $D^+_s\to K^+K^+\pi^-\pi^0$, if there are multiple $\pi^0$s
on the signal side, only the $\pi^0$ with the smallest $\chi^2$ of the mass-constrained kinematic fit is selected.

\begin{figure*}[htbp]
  \centering
\includegraphics[width=1.0\textwidth]{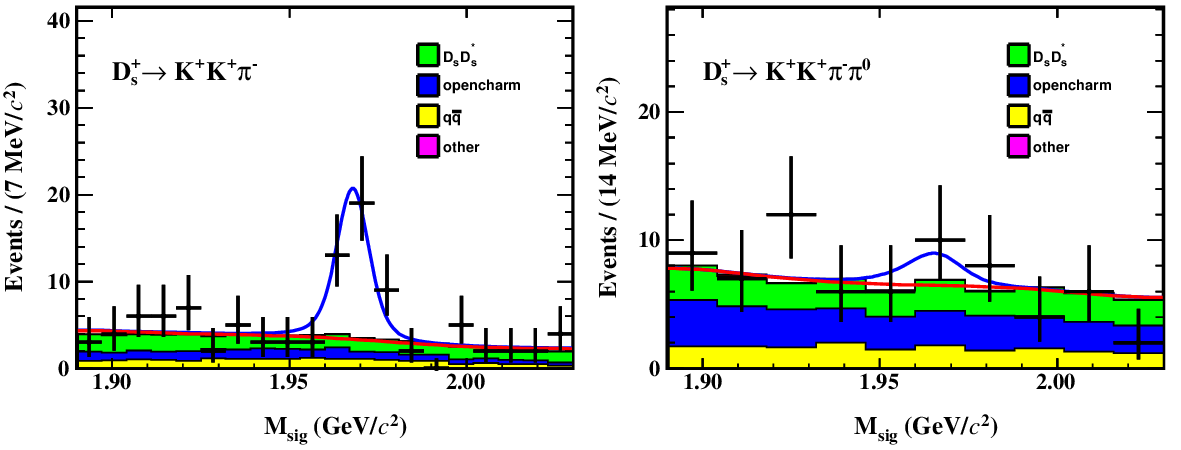}
  \caption{Fits to the $M_{\rm sig}$ distributions of the accepted DT candidates.
  The points with error bars are the combined data from all energy points. The blue solid curves are the fit results, the red solid curves are the combinatorial backgrounds, and the histograms filled in colors are from different background sources, derived from the inclusive MC sample and normalized to the fitted background yields in data.}
\label{fig:fit_Ntag}
\end{figure*}

Figure~\ref{fig:fit_Ntag} shows the distributions of the invariant masses, $M_{\rm sig}$, of the accepted signal candidates. The DT efficiencies and the signal efficiencies of detecting $D^+_s\to K^+K^+\pi^-$ and $D^+_s\to K^+K^+\pi^-\pi^0$ for different tag modes, weighted over different energy points, are summarized in Table 2. The average signal efficiencies of $D^+_s\to K^+K^+\pi^-$ and $D^+_s\to K^+K^+\pi^-\pi^0$, weighted over different tag modes, are
$(35.57\pm0.18)\%$ and $(10.32\pm0.10)\%$, respectively, where the uncertainties are due to the limited MC statistics.

The signal yields are extracted from unbinned maximum likelihood fits
to the $M_{\rm sig}$ spectra in Fig.~\ref{fig:fit_Ntag}. In the fit,
the signal is modeled by the MC-simulated signal shape smeared with
an additional Gaussian resolution function (with parameters obtained
from a fit to the corresponding CF decays).
The background is described by the simulated shape from the inclusive
MC sample, and the yields of the signal and background are floating.
The statistical significance of $D^+_s\to K^+K^+\pi^-$ is estimated to be $6.2\sigma$,
by $\sqrt{-2{\rm ln ({\mathcal L_0}/{\mathcal
      L_{\rm max}}})}$, where
${\mathcal L}_{\rm max}$ and ${\mathcal L}_0$ are the maximal
likelihoods of the fits with and without the signal contribution, respectively.
The signal yield of $D^+_s\to K^+K^+\pi^-$ is obtained to be
$32.9^{+7.6}_{-6.8}$.

For $D^+_s\to K^+ K^+ \pi^-\pi^0$, no significant signal is observed,
and an upper limit on the decay branching fraction is set.

\begin{figure}[htbp]
  \centering
\includegraphics[width=0.5\textwidth]{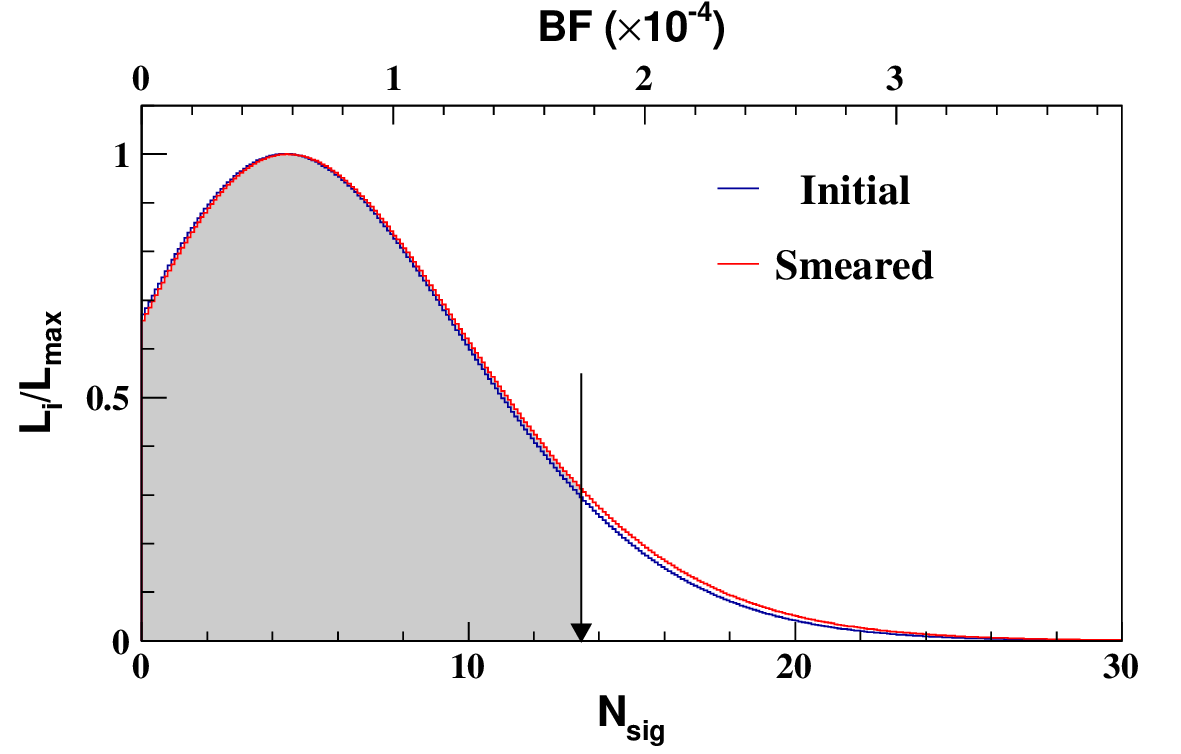}
  \caption{Distributions of normalized likelihoods versus the branching fraction of $D^+_s\to K^+K^+\pi^-\pi^0$.
The results incorporating the systematic uncertainties are shown as the blue curve. The black arrow shows the $N_{\rm sig}$ result corresponding to 90\% confidence level.}
\label{fig:upper}
\end{figure}

\section{Systematic uncertainties}

Table~\ref{tab:totsys} summarizes the sources of systematic uncertainty
in the branching fraction measurements; details are given below.

The systematic uncertainties on the fitted yields of the ST $D^-_s$ mesons are estimated by
using alternative signal and background shapes.
First, alternative signal shapes are obtained by changing from those
derived from the inclusive MC sample to those from the signal MC samples.
Second, an alternative background shape is obtained by varying the order
of the nominal Chebychev function by $\pm 1$.
For a given ST mode, the differences in the ratio of the yields of ST $D^-_s$ mesons over the corresponding efficiency for all variations,
and the background fluctuation of the fitted yield of ST $D^-_s$ are re-weighted by the yields of ST $D^-_s$ mesons in various data samples and are added in quadrature.  An additional component to this uncertainty is statistical in nature, and accounts for the contribution of background fluctuations to the fitted yields of ST $D^-_s$ mesons.
The total uncertainty associated with the ST yield $N_{\rm ST}^{\rm tot}$ is estimated to be 0.5\%.

The systematic uncertainties from the tracking and PID of $K^\pm$ and $\pi^\pm$ are estimated by using the control sample of
$e^+e^-\to K^+K^-\pi^+\pi^-(\pi^0)$~\cite{luyu}.
The systematic uncertainties in the tracking and PID efficiencies are both assigned as 1.0\% per $K^\pm$ or $\pi^\pm$.
The systematic uncertainties in the photon selection
and the $\pi^0$ reconstruction are studied with the control sample of $J/\psi\to\pi^+\pi^-\pi^0$~\cite{gammapipijpsi}.
The systematic uncertainty in the photon selection is assigned as 1.0\% per photon, and
the systematic uncertainty in the $\pi^0$ reconstruction, including the photon finding algorithm, the $\pi^0$ mass window and the 1C kinematic fit, is assigned as 2.0\% per $\pi^0$.

The systematic uncertainty related to the $M_{\rm sig}$ fit has two sources.
Firstly, an alternative signal shape is obtained by varying the parameters of the smeared Gaussian function by $\pm 1\sigma$.
Secondly, an alternative background shape is obtained by varying the $q\bar q$ component in the inclusive MC sample
by its uncertainty due to the limited MC statistics. For $D^+_s\to K^+K^+\pi^-$, the systematic uncertainty is evaluated as 0.4\%.
For the upper limit determination of $D^+_s\to K^+K^+\pi^-\pi^0$, the signal shape and background shape affect the likelihood function directly.
Changing the signal (background) shape shifts the signal yield from the fit by
0.1 (0.2), and we take the larger value as the additive systematic uncertainty.

Varying the $D_s^*$ branching fractions~\cite{pdg2022} by $\pm 1\sigma$
changes the efficiency by 0.3\%; this is taken as the associated uncertainty.

The systematic uncertainty related to the MC model is estimated using alternative signal MC samples.
For $D^+_s\to K^+K^+\pi^-$, the branching fraction of the $D^+_s\to K^+K^{*0}$ sub-channel is varied by its uncertainty.
For $D^+_s\to K^+K^+\pi^-\pi^0$, the average signal efficiency of $D^+_s\to K^{*0}K^+\pi^0$, $K^{*+}K^+\pi^-$, $K^+K^+\rho^-$, and $D^+_s\to K^+K^+\pi^-\pi^0$ (phase space) is used in place of the phase space efficiency.
The maximum changes of the signal efficiencies, 0.7\% and 5.3\%, are assigned as the systematic uncertainties for $D^+_s\to K^+K^+\pi^-$ and $D^+_s\to K^+K^+\pi^-\pi^0$, respectively.

\begin{table}[htpb]
\centering
\caption{Systematic uncertainties on the branching fraction measurements.
Uncertainties in the square brackets are additive uncertainties
on the number of events; all other uncertainties are in \%.}
\label{tab:totsys}
\begin{tabular}{c|ccc}
\hline\hline
Source            &   $K^+K^+\pi^-$  & $K^+K^+\pi^-\pi^0$    \\
\hline
$N_{\rm ST}$     &   0.5                    & 0.5    \\
Tracking          &   3.0                     & 3.0   \\
PID               &   3.0                     & 3.0    \\
$\gamma, \pi^{0}$ reconstruction &    1.0         & 3.0     \\
$M_{\rm sig}$ fit &   0.4              &  [0.2]    \\
Quoted branching fractions        &    0.3                      &   0.3      \\
MC model          &   0.7         &    5.3           \\
MC statistics     &   0.5         &    1.0           \\ \hline
Total             &   4.5         &    7.5 [0.2]  \\
\hline\hline
\end{tabular}%
\end{table}

\begin{table*}[htbp]
\centering
\caption{The branching fraction and upper limit obtained in this work,
the world average branching fractions of the corresponding CF decays,
the DCS/CF branching fraction ratios and these ratios in units of $\tan^{4}{\theta}_{\rm C}$.}
\begin{tabular}{c|c||c|c||c|c}
\hline\hline
DCS decay & ${\mathcal B}^{\rm this~work}_{\rm DCS}~(\times 10^{-4})$     & CF decay & ${\mathcal B}^{\rm PDG}_{\rm CF}~(\times 10^{-2})$    & ${\mathcal B}^{\rm this~work}_{\rm DCS}/{\mathcal B}^{\rm PDG}_{\rm CF}~(\times 10^{-3})$       & $\times \tan^{4}{\theta}_{\rm C}$      \\ \hline
$D_s^+\to K^+K^+\pi^-$           & ${1.23^{+0.28}_{-0.25}}\pm0.06$ & $D_s^+\to K^+K^-\pi^+$            & $5.37\pm0.10$ & $2.28^{+0.52}_{-0.47}$ & $0.79^{+0.18}_{-0.16}$ \\
$D_s^+\to K^+K^+\pi^-\pi^0$      & $<1.7$ & $D_s^+\to K^+K^-\pi^+\pi^0$        & $5.50\pm0.24$  & $<3.09$    &$<1.07$ \\ \hline\hline
\end{tabular}%
\label{totresult}
\end{table*}

\section{Results}

Inserting the numbers of $N_{\rm DT}$, $N_{\rm ST}$ and $\epsilon_{\rm sig}$ of
$D^+_s\to K^+K^+\pi^-$ in Eq.~(\ref{eq1}), we determine the decay branching fraction to be
\begin{eqnarray}
{\mathcal B}_{D^+_s\to K^+K^+\pi^-}=({1.23^{+0.28}_{-0.25}}({\rm stat})\pm0.06({\rm syst}))\times 10^{-4}.\nonumber
\end{eqnarray}

The upper limit on the branching fraction of $D^+_s\to K^+ K^+ \pi^-\pi^0$ is set with the Bayesian approach~\cite{K.Stenson:2006}, incorporating the systematic uncertainties. The raw likelihood distribution versus the branching fractions is smeared by a Gaussian function with a mean of zero and a width equal to the systematic uncertainty. The red solid curve in Fig.~\ref{fig:upper} shows the resulting likelihood distribution.
The upper limit on the branching fraction at 90\% confidence level is set as
\begin{eqnarray}
{\mathcal B}_{D^+_s\to K^+K^+\pi^-\pi^0}<1.7\times 10^{-4}. \nonumber
\end{eqnarray}

\section{Summary}

Using 7.33~fb$^{-1}$  of $e^+e^-$ collision data
collected at $E_{\rm cm}$ between 4.128 and 4.226 GeV with the BESIII detector,
we investigate the DCS decays $D^+_s\to K^+K^+\pi^-$ and $D^+_s\to K^+K^+\pi^-\pi^0$.
The absolute branching fraction of $D^+_s\to K^+K^+\pi^-$ is determined to be $({1.23^{+0.28}_{-0.25}}({\rm stat})\pm0.06({\rm syst}))\times 10^{-4}$,
which is in good agreement with the world average value of $(1.274 \pm 0.031)\times 10^{-4}$~\cite{pdg2022},
based on the {\it relative} measurements from LHCb, BaBar, Belle and
FOCUS~\cite{pap34,pap35,pap36,pap37}.
No significant signal of $D^+_s\to K^+K^+\pi^-\pi^0$ is observed.
The upper limit on the branching fraction of $D^+_s\to K^+K^+\pi^-\pi^0$
is $1.7\times10^{-4}$ at 90\% confidence level.
Table \ref{totresult} summarizes the results obtained in this work and
the world average branching fractions of the corresponding CF decays.
The DCS/CF branching fraction ratios and the corresponding factors relative to $\tan^{4}{\theta}_{\rm C}$
are also listed. No significant deviation from native expectation of $(0.5-2.0)\times {\rm tan}^4\theta_C$ is found.

\section{Acknowledgement}

The BESIII Collaboration thanks the staff of BEPCII and the IHEP computing center for their strong support. This work is supported in part by National Key R\&D Program of China under Contracts Nos. 2020YFA0406400, 2020YFA0406300; National Natural Science Foundation of China (NSFC) under Contracts Nos. 11635010, 11735014, 11835012, 11935015, 11935016, 11935018, 11961141012, 12025502, 12035009, 12035013, 12061131003, 12192260, 12192261, 12192262, 12192263, 12192264, 12192265, 12221005, 12225509, 12235017; the Chinese Academy of Sciences (CAS) Large-Scale Scientific Facility Program; the CAS Center for Excellence in Particle Physics (CCEPP); Joint Large-Scale Scientific Facility Funds of the NSFC and CAS under Contract No. U1932102, U1832207; CAS Key Research Program of Frontier Sciences under Contracts Nos. QYZDJ-SSW-SLH003, QYZDJ-SSW-SLH040; 100 Talents Program of CAS; The Institute of Nuclear and Particle Physics (INPAC) and Shanghai Key Laboratory for Particle Physics and Cosmology; European Union's Horizon 2020 research and innovation programme under Marie Sklodowska-Curie grant agreement under Contract No. 894790; German Research Foundation DFG under Contracts Nos. 455635585, Collaborative Research Center CRC 1044, FOR5327, GRK 2149; Istituto Nazionale di Fisica Nucleare, Italy; Ministry of Development of Turkey under Contract No. DPT2006K-120470; National Research Foundation of Korea under Contract No. NRF-2022R1A2C1092335; National Science and Technology fund of Mongolia; National Science Research and Innovation Fund (NSRF) via the Program Management Unit for Human Resources \& Institutional Development, Research and Innovation of Thailand under Contract No. B16F640076; Polish National Science Centre under Contract No. 2019/35/O/ST2/02907; The Swedish Research Council; U. S. Department of Energy under Contract No. DE-FG02-05ER41374.


\begin{thebibliography}{99}

\bibitem{theory_1}
H. Y. Cheng and C. W. Chiang,
\href{https://journals.aps.org/prd/abstract/10.1103/PhysRevD.81.074021}{ Phys. Rev. D {\bf 81}, 074021 (2010).}

\bibitem{Lipkin}
H. J. Lipkin,
\href{https://www.sciencedirect.com/science/article/pii/S0920563202019655?via\%3Dihub}{Nucl. Phys. Suppl. {\bf 115}, 117 (2003).}

\bibitem{bes3_DCS_Kpipipi0}
M. Ablikim {\it et al.} (BESIII Collaboration),
\href{https://journals.aps.org/prl/abstract/10.1103/PhysRevLett.125.141802}{Phys. Rev. Lett. {\bf 125}, 141802 (2020).}

\bibitem{bes3-DCS-Dp-K3pi-v2}
M.~Ablikim {\it et al.} (BESIII Collaboration),
\href{http://link.aps.org/doi/10.1103/PhysRevD.104.072005}{Phys. Rev. D {\bf 104}, 072005 (2021).}

\bibitem{pap4}
Z.~z.~Xing,
\href{https://doi.org/10.1142/S0217732319502389}{Mod. Phys. Lett. A \textbf{34}, 1950238 (2019).}


\bibitem{pap5}
H.~J.~Lipkin,
\href{https://doi.org/10.1103/PhysRevLett.46.1307}{Phys. Rev. Lett. \textbf{46}, 1307 (1981).}


\bibitem{pap6}
Q.~Qin, H.~n.~Li, C.~D.~L\"u and F.~S.~Yu,
\href{https://doi.org/10.1103/PhysRevD.89.054006}{Phys. Rev. D \textbf{89}, 054006 (1981).}


\bibitem{pap7}
H.~Y.~Cheng, C.~W.~Chiang and A.~L.~Kuo,
\href{https://doi.org/10.1103/PhysRevD.93.114010}{Phys. Rev. D \textbf{93}, 114010 (2016).}


\bibitem{pap8}
W.~Kwong and S.~P.~Rosen,
\href{https://doi.org/10.1016/0370-2693(93)91843-C}{Phys. Lett. B \textbf{298}, 413 (1993).}


\bibitem{pap9}
Y.~Grossman and D.~J.~Robinson,
\href{https://doi.org/10.1007/JHEP04(2013)067}{J. High Energy Phys. \textbf{04}, 067 (2013).}


\bibitem{pap10}
H.~n.~Li, C.~D.~L\"u and F.~S.~Yu,
\href{https://doi.org/10.1103/PhysRevD.86.036012}{Phys. Rev. D \textbf{86}, 036012 (2012).}



\bibitem{pdg2022}
	P. A. Zyla {\it et al.} (Particle Data Group),
	\href{https://pdg.lbl.gov/}{Prog. Theor. Exp. Phys. {\bf 2022}, 083C01 (2022).}

\bibitem{pap34}
R.~Aaij \textit{et al}. (LHCb Collaboration),
\href{https://inspirehep.net/literature/1697372}{JHEP \textbf{03}, 176 (2019).}

\bibitem{pap35}
P.~del Amo Sanchez \textit{et al}. (BaBar Collaboration),
\href{https://inspirehep.net/literature/878120}{Phys. Rev. D \textbf{83}, 052001 (2011).}

\bibitem{pap36}
B.~R.~Ko \textit{et al}. (Belle Collaboration),
\href{https://inspirehep.net/literature/816676}{Phys. Rev. Lett. \textbf{102}, 221802 (2009).}

\bibitem{pap37}
J.~M.~Link \textit{et al}. (FOCUS Collaboration),
\href{https://inspirehep.net/literature/688341}{Phys. Lett. B \textbf{624}, 166 (2005).}

\bibitem{BESIII}
M. Ablikim {\it et al.} (BESIII Collaboration),
\href{https://doi.org/10.1016/j.nima.2009.12.050}{Nucl. Instrum. Meth. A {\bf 614}, 345 (2010).}

\bibitem{Yu:IPAC2016-TUYA01}
C.~H.~Yu {\it et al.},
\href{doi:10.18429/JACoW-IPAC2016-TUYA01}{\it Proceedings of IPAC2016, Busan, Korea, (2016).}

\bibitem{Pcao}
P.~Cao {\it et al.},
\href{https://doi.org/10.1016/j.nima.2019.163053}{Nucl. Instrum. Meth. A {\bf 953}, 163053 (2020).}

\bibitem{dt1}
M. Ablikim {\it et al.} (BESIII Collaboration),
\href{http://dx.doi.org/10.1088/1674-1137/40/6/063001}{Chin. Phys. C {\bf 40}, 063001 (2016).}

\bibitem{dt2}
M. Ablikim {\it et al.} (BESIII Collaboration),
\href{http://dx.doi.org/10.1088/1674-1137/ac1575}{Chin. Phys. C {\bf 45}, 103001 (2020).}

\bibitem{geant4}
S. Agostinelli {\it et al.} (GEANT4 Collaboration),
\href{https://doi.org/10.1016/S0168-9002(03)01368-8}{Nucl. Instrum. Meth. A {\bf 506}, 250 (2003).}

\bibitem{kkmc}
S. Jadach, B. F. L. Ward, and Z. Was,
\href{https://linkinghub.elsevier.com/retrieve/pii/S0010465500000485}{ Comp. Phys. Commu. {\bf 130}, 260 (2000);} \href{https://journals.aps.org/prd/abstract/10.1103/PhysRevD.63.113009}{Phys. Rev. D {\bf 63}, 113009 (2001).}

\bibitem{evtgen}
D.~J.~Lange,
\href{https://doi.org/10.1016/S0168-9002(01)00089-4} {Nucl. Instrum. Meth. A {\bf 462}, 152 (2001);}
R.~G.~Ping,
\href{https://doi.org/10.1088/1674-1137/32/8/001}{Chin. Phys. C {\bf 32}, 599 (2008).}

\bibitem{lundcharm}
J. C. Chen, G. S. Huang, X. R. Qi, D. H. Zhang, and Y. S. Zhu,
\href{https://journals.aps.org/prd/abstract/10.1103/PhysRevD.62.034003}{Phys. Rev. D {\bf 62}, 034003 (2000).}

\bibitem{photos}
E.~Richter-Was,
\href{https://doi.org/10.1016/0370-2693(93)90062-M`}{Phys. Lett. B {\bf 303}, 163 (1993).}


\bibitem{DTmethod}
R. M. Baltrusaitis {\it et al.} (MARKIII Collaboration),
\href{https://journals.aps.org/prl/abstract/10.1103/PhysRevLett.56.2140}{Phys. Rev. Lett. {\bf 56}, 2140 (1986);}
J. Adler {\it et al.} (MARKIII Collaboration),
\href{https://journals.aps.org/prl/abstract/10.1103/PhysRevLett.60.89}{Phys. Rev. Lett. {\bf 60}, 89 (1988).}

\bibitem{bes2019}
M. Ablikim {\it et al.} (BESIII Collaboration),
\href{https://journals.aps.org/prl/abstract/10.1103/PhysRevLett.122.071802}{Phys. Rev. Lett. {\bf 122}, 071802 (2019).}

\bibitem{bes3_etaev}
M. Ablikim {\it et al.} (BESIII Collaboration),
\href{https://journals.aps.org/prl/abstract/10.1103/PhysRevLett.122.121801}{Phys. Rev. Lett. {\bf 122}, 121801 (2019).}

\bibitem{bes3_gev}
M. Ablikim {\it et al.} (BESIII Collaboration),
\href{https://journals.aps.org/prd/pdf/10.1103/PhysRevD.99.072002}{Phys. Rev. D {\bf 99}, 072002 (2019).}


\bibitem{luyu}
M. Ablikim {\it et al.} (BESIII Collaboration),
\href{https://link.aps.org/doi/10.1103/PhysRevD.99.091101} {Phys. Rev. D \textbf{99}, 091101 (2019).}



\bibitem{gammapipijpsi}
M. Ablikim \textit{et al}. (BESIII Collaboration),
\href{https://journals.aps.org/prd/abstract/10.1103/PhysRevD.83.112005} {Phys. Rev. D \textbf{83}, 112005 (2011).}



\bibitem{K.Stenson:2006}
K.~Stenson,
\href{https://arxiv.org/abs/physics/0605236}{arXiv:0605236[physics].}




\end{thebibliography}
\end{document}